# Decoding the mechanisms of phase transitions from *in situ* microscopy observations

Mani Valleti,[1,2] Reinis Ignatans,[3] Sergei V. Kalinin,[2,a] and Vasiliki Tileli[3,b]

[1] Bredesen Center for Interdisciplinary Research, University of Tennessee, Knoxville, TN 37996, USA

[2] The Center for Nanophase Materials Sciences, Oak Ridge National Laboratory, Oak Ridge, TN 37831, USA

[3] Institute of Materials, École polytechnique fédérale de Lausanne, Station 12, 1015 Lausanne, Switzerland

Temperature-induced phase transition in $BaTiO_3$ has been explored using the machine learning analysis of domain morphologies visualized via variable-temperature scanning transmission electron microscopy (STEM) imaging data. This approach is based on the multivariate statistical analysis of the time or temperature dependence of the statistical descriptors of the system, derived in turn from the categorical classification of observed domain structures or projection on the continuous parameter space of the feature extraction-dimensionality reduction transform. The proposed workflow offers a powerful tool for exploration of the dynamic data based on the statistics of image representation as a function of external control variable to visualize the transformation pathways during phase transitions and chemical reacitons. This can include the mesoscopic STEM data as demonstrated here, but also optical, chemical imaging, etc. data. It can further be extended to the higher dimensional spaces, for example analysis of the combinatorial libraries of materials compositions.

[a] sergei2@ornl.gov
[b] vasiliki.tileli@epfl.ch

Phase transitions are keystone across multiple areas of condensed matter physics and materials science and underpin broad variety of the manufacturing and fabrication processes. Ferroelectric and magnetic phase transitions are broadly utilized in information technology devices, including current and emergent non-volatile memories and logic devices, as well as data storage.[1,2] Structural phase transitions in solids are involved virtually all areas of materials processing from metallurgy to ceramics and construction. Finally, phenomena such as intercalation, alloying, and unmixing during chemical transformations at surfaces and interfaces associated with the formation of spatially non-uniform patterns and microstructures play the key role in energy technologies including batteries and fuel cells, catalysis, and many others.[3-5]

These broad variety of phase transformation and the role they play in applications have long made them one of the principle targets for fundamental research. From the classical condensed matter physics perspective, the universal framework for description of phase transition can be derived based on the symmetries that break during the phase transition. For ferroelectricity, ferromagnetism, and superconductivity the natural framework for description of phase transition is given by the Ginzburg-Landau theory,[2] in which the form of the order parameter is intrinsically linked to the symmetry of the underpinning lattice and the property of interest. This approach allows description of the phase transition, analysis of the associated changes in materials functionalities, and prediction of the possible topological defects in low-symmetry phases. Alternatively, lattice models can be used to analyzed phase transition phenomena, as exemplified by a range of possible models ranging from by now textbook Ising model to modern lattice Hamiltonian theories.[6]

From experimental perspective, the phase transitions are traditionally explored using combination of scattering techniques and macroscopic property measurements. Here, functional responses such as optical properties, permittivity, susceptibility, or magnetization provide the information on the phase transition temperature and character. The scattering data provides information on the associated symmetry changes and descriptors such as lattice parameters that are strongly coupled to the order parameter. However, this approach is limited when applied to materials with poorly defined ground states that lack discrete translational symmetries, including structural and physical glasses,[7,8] systems such as nanoscale phase separated materials,[6,9,10] ferroelectric relaxors[11-13] and morphotropic phase boundary[14-16] systems. This limitation is particularly significant given that these materials tend to exhibit giant functional responses such as high susceptibilities or electromechanical couplings across broad temperature or field ranges, making them ideal for multiple applications. Hence, understanding of phase transitions in such materials represents an obvious fundamental and applied interest.

Recently, the advances in imaging techniques including *in situ* electron and atomic force microscopy allow to observe dynamic phenomena, ranging from formation of the topological defects and phase transformation fronts,[17-19] formations of ferroelectric and phase ordered domains,[20] chemical segregation and reactions,[21,22] to name a few. Recent advances in atomically resolved imaging allowed atomic-scale observations of thermal and electron beam induced transformations on the atomic level.[23-26] Similarly, optical and scanning probe microscopy techniques can be used for observation of discrete systems such as particle self-organization on the solid-liquid interfaces and colloid crystals.[27, 28] Both for the mesoscopic and discrete systems, the observations can be performed as a function of external control stimuli including temperature and electric or magnetic fields, visualizing the evolution of microstructure as a function of these control variables over time. This naturally leads to the question of whether these observations can be used to establish the presence and quantify the dynamics of the phase transitions from observational data only.

While simple in concept, realization of this approach requires several issues to be addressed. First, the phase transition has to be associated with the emergence of statistically different populations of the observable defects and structural elements providing meso- or atomic scale identifiers. Secondly, the possible defect structures emerging due to intrinsic materials behavior, macroscopic and mesoscale field gradients, and interactions with frozen disorder should allow description in terms of small number of descriptor variables, equivalent to order parameter components in macroscopic theories.

Here, we develop such approach based on the Gaussian mixture models (GMM) that offers a framework for statistical clustering of the imaging data, separating into discrete classes. We further extend this analysis to the feature engineering transformation-dimensionality reduction workflow that allows projecting salient elements of domain structure onto the continuous parameter space, and explore the dynamics as a function of time or stimulus.

For implementing the algorithm, we use series of *in-situ* scanning transmission electron microscopy (STEM) images acquired in bright field (BF) mode of temperature-induced transformations in single crystalline $BaTiO_3$, a classical ferroelectric system. In particular, we follow the tetragonal (room temperature, RT) to cubic (over the Curie temperature, $T_c$) phase transition while monitoring the domain features. The specimen was focused ion beam prepared in a lamella configuration that was placed on MEMS-based chips for the real-time temperature-induced observations as described in detail elsewhere.[29] The sequence of transformations upon heating are shown in Fig. 1. At RT the sample is dominated by 180° domain walls, which is characteristic of geometry-restricted ferroelectric systems.[30] Upon heating, the domains transition to a morphology containing only 90° domain walls[31] (starting at 50 °C). Finally, the paraelectric state is observed above Tc (associated with loss of contrast from the domain walls). When the sample is cooled down, the phase transition temperature is measured at around 137 °C, which is signified by the appearance of 90° domain walls. Finally, the sample recovers the 180° domain wall initial state with some hysteresis, for example compare Fig. 1a with 1f. The morphologic transformation from 90° to 180° takes place at around 30 °C. It should be noted that the data used in this paper is also a part of dataset reported earlier[29, 32].

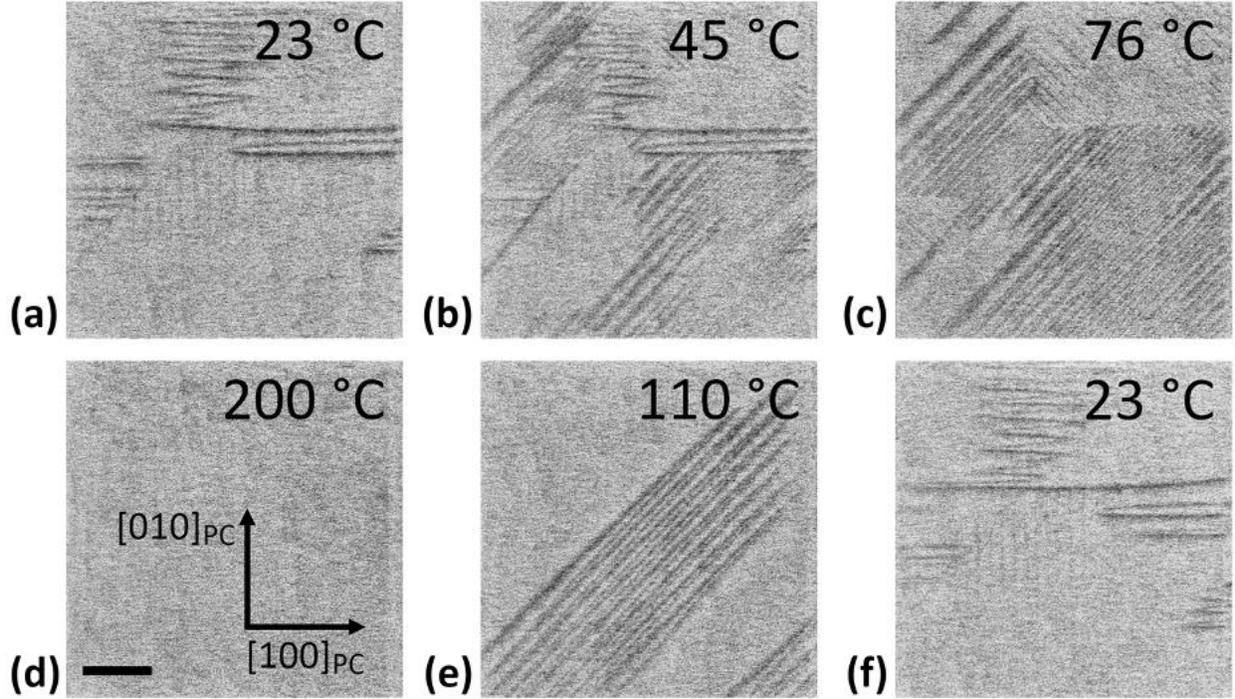

**Figure 1.** Domain evolution in the BaTiO3 lamella observed via BF-STEM as different temperatures during the heating and cooling cycle. The scale bar is 500 nm.

To determine the domain dynamics and infer phase transitions based on image analysis, we develop the machine learning image analysis workflow. Here, the stack of experimental images is partitioned into sub-images assuming continuous translational symmetry, i.e. the centers of the sub-images are chosen randomly. Note that while this approach is simple from the point of view of forming the feature set, it offers largest challenges from the perspective of physical interpretability, since the features describing materials behavior such as domain walls can have arbitrary shifts and orientations with respect to the sub-image center. For example, the continuous rotational symmetry has long been the hurdle in analysis of the high-resolution electron microscopy data in the systems undergoing chemical transformations, since the presence of multiple rotational variants precludes applicability of linear unmixing methods such as principal component analysis and non-linear matrix factorization.

Here, we explore and compare the applicability of two approaches, based on the fixed discrete, offset-invariant, and rotationally invariant representations of the system. The former approach is realized via the Gaussian mixture modelling, whereas the second approach is realized via feature engineering-dimensionality reduction transform.

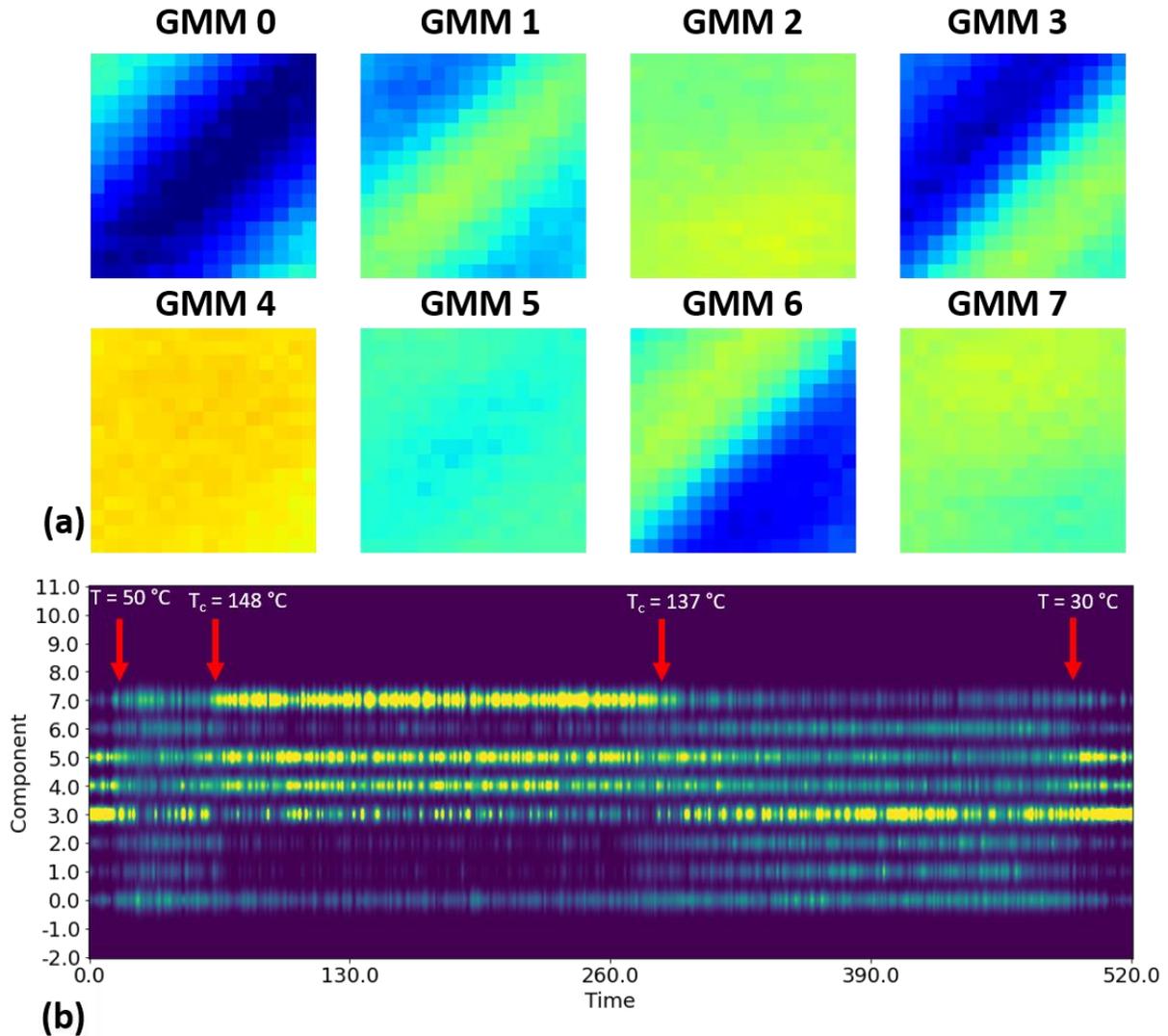

**Figure 2.** (a) GMM components for decomposition in 8 components. The vertical scale is (0.5, 0.8) for all components. (b) Time dependence of kernel density plot of the components. Note the rapid changes of GMM populations with the morphological transitions in domain population.

As a first approach, we have applied the GMM separation on the sub-images. GMM is a generalization of K-means clustering algorithm where each cluster is assumed to be derived from Gaussian distributions with unknown parameters[33]. They incorporate the information about the covariance structure of data and means of the clusters. From the 521 frames down sampled to 512x512 pixels, we extract 100 sub-images of size $n \times n$ with randomly chosen centers to yield 52k feature training set. The sub-image sizes were chosen as $n = 16$ and 64, however other image sizes were explored. Note that given the image size and sub-image sizes, the number of sub-images is close to the full non-overlapping tiling. The sub-image stack was used to train the GMM, yielding the GMM class for each sub-image.

The resultant means of GMM classes are shown in Fig. 2 (a). Note that in general these classes represent the features having the general orientation of the domain wall in the images, shifted in the directions of one of the primary diagonals. The population of the classes depends on the domain populations in the original image stack, and in this case the (1-1) direction was preponderant. Some of the classes are featureless and correspond to the part of the surface that does not have domain walls in the field of view. Upon increase of the number of GMM components, the classes corresponding to the orthogonal domain walls and 180 walls start to appear; however, the effective number of classes becomes very large. Furthermore, some of the classes are similar morphologically, and while more complex compound classes can be created by supervised or unsupervised combination of initial classes, this approach requires a large degree of human intervention. With these caveats, GMM allows separating the observed morphologies into specific classes.

Once the GMM model is trained, it can be used to map the evolution of the system as a function of external parameter. To achieve this, we calculate the number of sub-images pertaining to each GMM class for each image in the stack. For convenience of visualization and comparison with subsequent analysis, we visualize the data as a kernel density estimate plot as shown in Fig. 2 (b). Here, position of the band corresponds to the number of the GMM component, and the intensity to the fraction of the component. Note that clear transitions (marked arrows) are visible, indicative of drastic changes in the character of the domain structure in the material at these temperatures. Further note the shot-noise like behavior in the GMM component intensities, which corresponds to the ambiguity of class assignment in close morphologies that can be affected by the noise present in images.

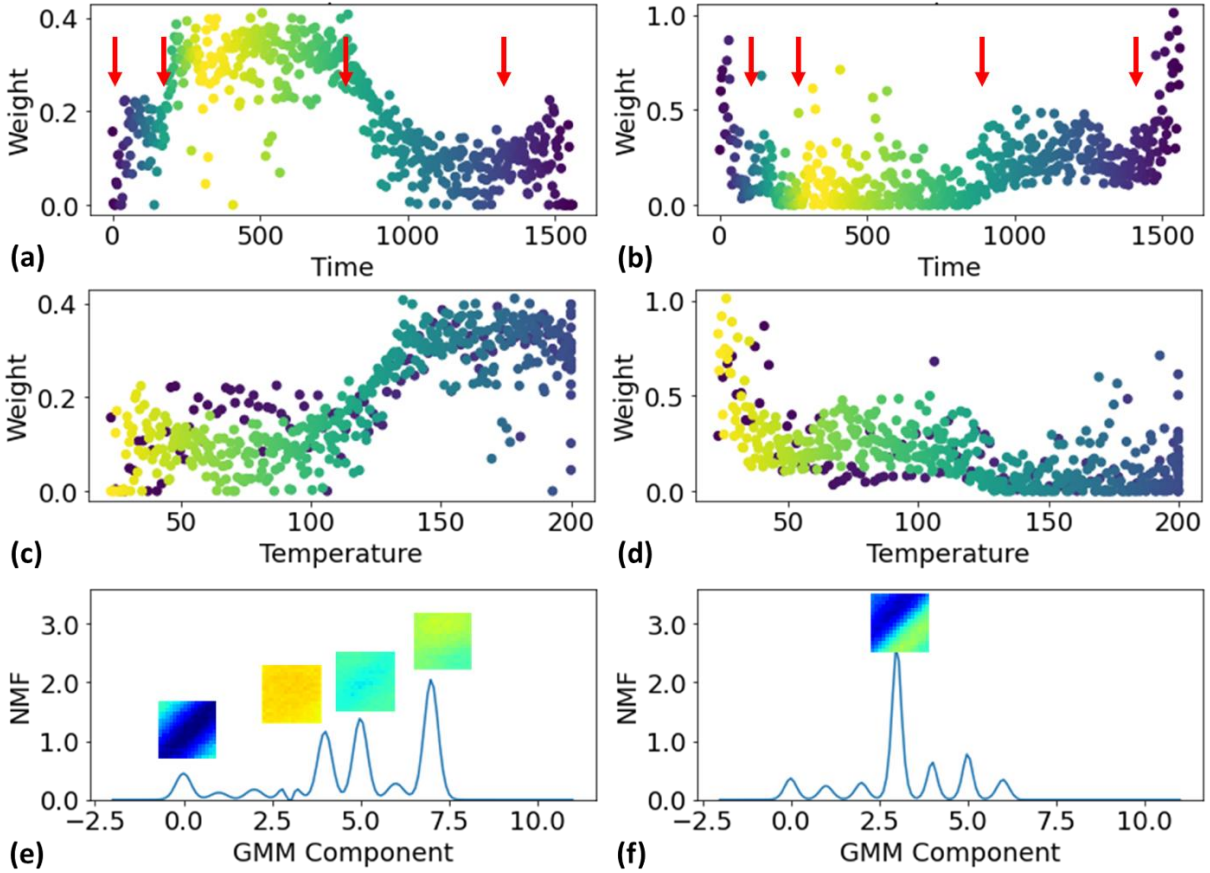

**Figure 3.** Evolution of the density of the NMF weights as a function of time and temperature. (a,b) Time dependence of the NMF weights for decomposition in 2 components. The color of the points corresponds to temperature. (c,d) The first NMF components as a function of temperature (°C). The color of the points corresponds to time. (e,f) The distribution of the GMM components in the NMF components and the corresponding GMM images. Note that of 8 GMM components, all 8 are present in the first component and 7 are present in second.

To get further insight in this behavior, the data were analyzed using the non-negative matrix factorization (NMF). We choose NMF as the simplest method that gives rise to non-negative components. In NMF, the time dependence of the kernel density estimate KDE(*c,t*), where *c* is a categorical variable corresponding to the GMM component and *t* is time, is represented as

$$KDE(c,t) = \sum_{i=1}^{N} L_i(c) w_i(t) \quad (1)$$

where $L_i(c)$ are the weights that represent the fraction of the GMM components in a specific domain morphology and $w_i(t)$ are the endmembers that determine characteristic time behaviors. The number of components *N* is set at the beginning of the analysis and can be chosen based on the quality of decomposition, anticipated physics of the systems, etc. Here, after experimentation, the $N = 2$ was found to be sufficient to represent the observed dynamics.

The time dependence of the 1st and 2nd NMF component are shown in Fig. 3 (a,b) respectively. Note that the time dynamics of the components representing the collective behavior of the domain morphologies shows clear rapid changes at the transition temperatures (red arrows corresponding to inflection points on the dependencies). This behavior can further be illustrated as a function of temperature as shown in Fig. 3 (c,d). Here, the first NMF component clearly shows the ferroelectric-paraelectric transition at 130 °C, whereas the second component shows both transitions at 30 and 130 °C.

The nature of the domain morphologies that change at these transitions can be illustrated via the NMF weights as shown in Fig. 3 (e,f). Here, the first component is dominated by weak or absent domain contrast, corresponding to predominantly paraelectric/noise dominated phase or the background parent domain. At the same time, the second component is dominated by the domain walls, corresponding to changes of the ferroelectric component (i.e. change of polarization direction). Note that in this description the NMF components do not comport to pure order parameters and rather represent statistical behaviors of domain populations. Nonetheless, this approach allows to determine the temperature and characteristics of phase transitions.

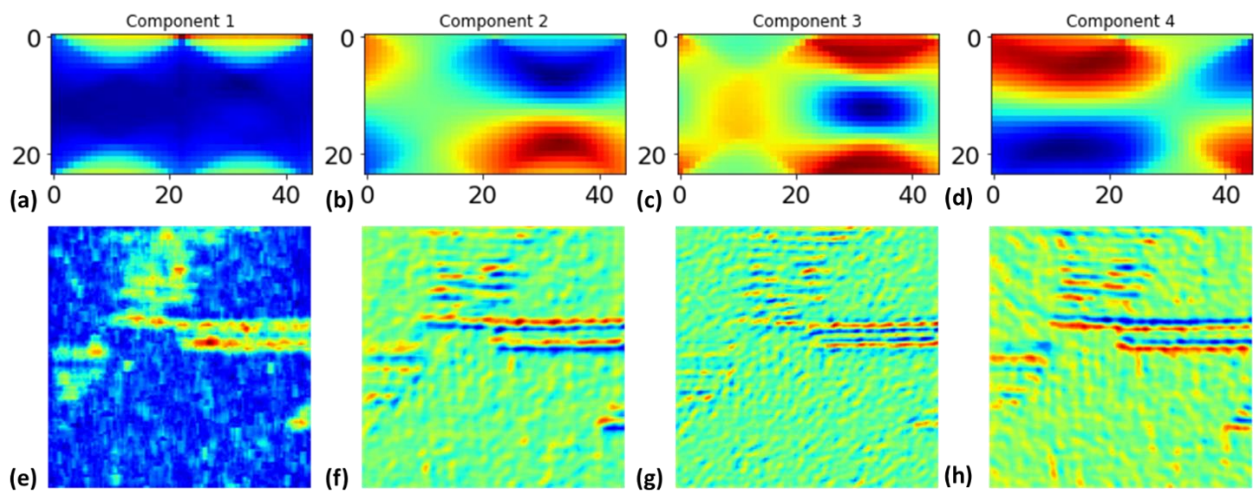

**Figure 4.** Principal component analysis of the Radon transforms of the sub-images. Shown are (a-d) first 4 PCA components of the Radon transforms, and (e-f) encoded image corresponding to Fig. 1 (a).

To further explore the evolution of the domain morphology, we develop the workflow based on the sliding window transform. Here, the stack of the experimental frames is parsed to generate several sub-images of predefined size centered at randomly chosen points within all the frames. Here, we have generated 500 sub-images of size 24x24 per frame, to the total of 260,500 sub-images. The features can be engineered from the sub-images using the appropriate transform, including identity, Fourier, or Radon. Subsequently, the feature set can be simplified using the linear or non-linear decomposition or clustering method. Here, we incorporated the principal component analysis, non-negative matrix factorization, and Gaussian mixture model as an option. The choice of the feature engineering and unmixing methods are determined by the characteristic features of the domain structure. The weighs of the transformation can then be explored as function of time, providing the insight into the temperature and time dependent behavior. Similarly, the trained transform (principal components) can be used to decode the features in individual images. In this, for each pixel in the image we generate a sub-image centered at the pixel, and then decode it using the previously trained transform. For the practical reasons, we implement the process with a certain stride, i.e. every *s*-th pixel is decoded. Here, *s* is taken as 4, well below the sub-image size. Note that compared to previously introduced sliding transform methods,[34-37] here the decoding of a single image is performed using the transform trained on the whole data set, and thus contains information pertinent to the overall physics of the process.

Here, we have chosen Radon transform for feature engineering, since it is ideally suited for the discovery of linear features such as domain walls. Shown in Fig. 4 (a-d) are the first 4 principal component analysis (PCA) components of the Radon transform defined on the whole sub-image stack. In comparison, shown in Fig. 4 (e-h) is the decoded image corresponding to frame 1. Note that the first PCA-Radon component encodes the domain walls, whereas higher order components exhibit differential contrast corresponding to domain wall roughness. In comparison, using the PCA-FFT analysis gives rise to a larger number of non-trivial components, since Fourier transform is less suited for discovery of linear patterns.

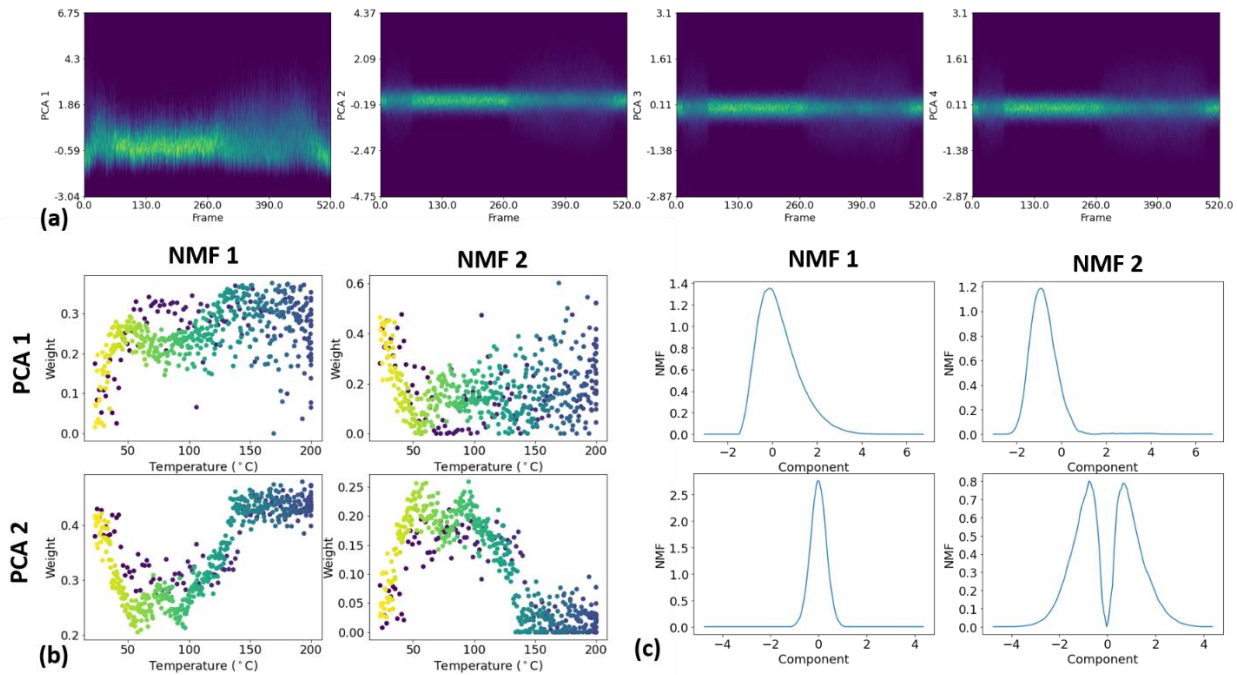

**Figure 5.** (a) Evolution of the kernel density plots for the PCA-Radon transform as a function of frame number. (b) The dependence of the NMF component weights for the first and second PCA-Radon transform as a function of temperature, color indicates the frame number,(c) The representations of the NMF components for the first two principal components.

The evolution of the decoded sub-images during the experiment can be used to infer the phase transitions. To quantify this behavior, we have transformed the distributions of the PCA weights for each time frame into the corresponding kernel density estimate, i.e. approximate probability distribution. The $KDE_i(c, time)$, where c is continuous range variable, time is the frame number, and *i* is the number of PCA-Radon component, is shown in Fig. 5 (a). Note the clear changes in the distributions at certain temperatures.

To simplify this behavior further, we decompose the KDE using the NMF. Here, NMF is chosen since the KDE components are by definition non-negative. Given that behavior of $2^{nd}$ and higher PCA-Radon component are very similar, we show the decomposition only for the first two components in Fig. 5 (b). Here, the components are plotted as a function of temperature, and clearly visualize phase transition regions and the hysteresis in domain structure.

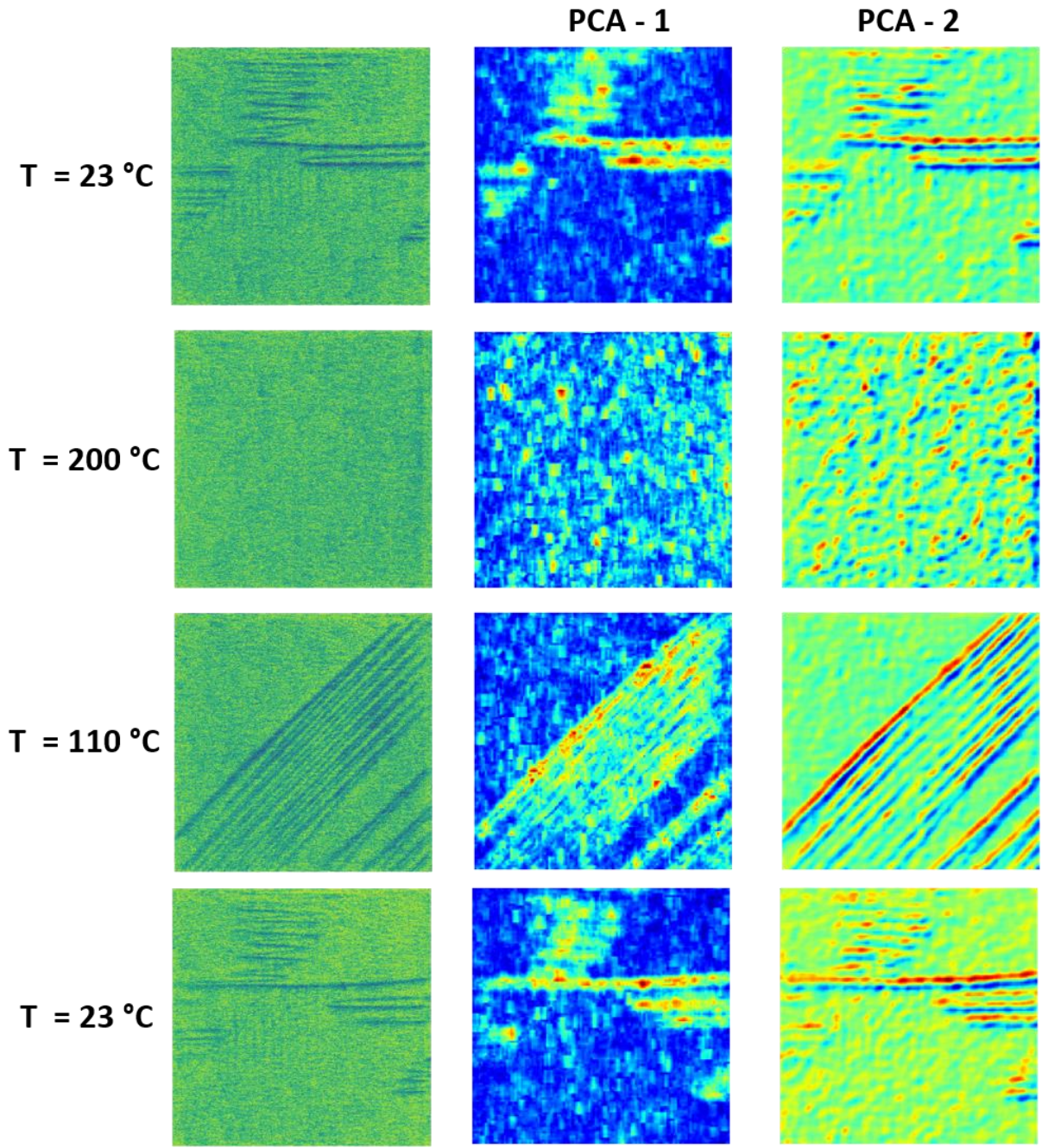

**Figure 6.** Evolution of decoded images using the first two principal components at different temperatures

Finally, the evolution of the decoded images is shown in Fig. 6. Here, the original images as well as spatial maps of the decoded transform components clearly illustrate the evolution of the domain structure. Note that the transformation has been trained on the full data set both before and above the transition, allowing for analysis of the structures above Tc.

To summarize, here we proposed and implemented the workflow for analysis of the variable-temperature imaging data to explore the phase transition from the changes in domain morphologies. This approach is based on the multivariate statistical analysis of the time or temperature dependence of the statistical descriptors of the system. These in turn can be derived from the categorical classification of observed domain structures, or continuous projection of the observed structures on the parameter space of the feature extraction-dimensionality reduction transform. This analysis does not require the spatial alignment, and can be extended to hyperspectral datasets.

More generally, the proposed workflow offers a powerful tool for exploration of the dynamic data based on the statistics of image representation as a function of external control variable to visualize the transformation pathways, phase transitions, etc. This can include the mesoscopic STEM data as demonstrated here, but also optical, chemical imaging, etc. data. It can further be extended to the higher dimensional stimulus spaces, for example analysis of the combinatorial libraries of materials compositions under static and dynamic (e.g. varying temperature) conditions.

**Acknowledgements:** This effort (machine learning) is based upon work supported by the U.S. Department of Energy (DOE), Office of Science, Basic Energy Sciences (BES), Materials Sciences and Engineering Division (S.M.V., S.V.K.) and was performed at the Oak Ridge National Laboratory's Center for Nanophase Materials Sciences (CNMS), a U.S. Department of Energy, Office of Science User Facility. R. I. and V. T. acknowledge financial support from the Swiss National Science Foundation (SNSF) under award no. 200021_175711. The authors gratefully acknowledge Maxim Ziatdinov (ORNL) for multiple useful discussions and advice.

# References

1. R. Waser, Nanoelectronics and Information Technology (2012).
2. L. E. C. A.K. Tagantsev, and J. Fousek, *Domains in Ferroic Crystals and Thin Films*. (Springer, New York, 2010).
3. J. B. Goodenough and Y. Kim, Chemistry of Materials **22** (3), 587-603 (2010).
4. A. Kraytsberg and Y. Ein-Eli, Journal of Power Sources **196** (3), 886-893 (2011).
5. J. M. Tarascon and M. Armand, Nature **414** (6861), 359-367 (2001).
6. E. Dagotto, Science **309** (5732), 257-262 (2005).
7. K. Binder and A. P. Young, Rev. Mod. Phys. **58** (4), 801-976 (1986).
8. K. Binder and J. D. Reger, Advances in Physics **41** (6), 547-627 (1992).
9. L. Samet, D. Imhoff, J. L. Maurice, J. P. Contour, A. Gloter, T. Manoubi, A. Fert and C. Colliex, European Physical Journal B **34** (2), 179-192 (2003).
10. E. Dagotto, T. Hotta and A. Moreo, Physics Reports-Review Section of Physics Letters **344** (1-3), 1-153 (2001).
11. A. E. Glazounov, A. K. Tagantsev and A. J. Bell, Phys. Rev. B **53** (17), 11281-11284 (1996).
12. V. Westphal, W. Kleemann and M. D. Glinchuk, Phys. Rev. Lett. **68** (6), 847-850 (1992).
13. B. E. Vugmeister, Phys. Rev. B **73** (17) (2006).
14. D. Damjanovic, Rep. Prog. Phys. **61** (9), 1267-1324 (1998).
15. I. Grinberg, M. R. Suchomel, P. K. Davies and A. M. Rappe, J. Appl. Phys. **98** (9) (2005).
16. D. I. Woodward, J. Knudsen and I. M. Reaney, Phys. Rev. B **72** (10) (2005).
17. Q. Li, C. T. Nelson, S. L. Hsu, A. R. Damodaran, L. L. Li, A. K. Yadav, M. McCarter, L. W. Martin, R. Ramesh and S. V. Kalinin, Nat. Commun. **8** (2017).
18. T. J. Pennycook, L. Jones, H. Pettersson, J. Coelho, M. Canavan, B. Mendoza-Sanchez, V. Nicolosi and P. D. Nellist, Sci Rep **4** (2014).
19. H. M. Zheng, J. B. Rivest, T. A. Miller, B. Sadtler, A. Lindenberg, M. F. Toney, L. W. Wang, C. Kisielowski and A. P. Alivisatos, Science **333** (6039), 206-209 (2011).
20. C. L. Jia, K. W. Urban, M. Alexe, D. Hesse and I. Vrejoiu, Science **331** (6023), 1420-1423 (2011).
21. M. R. Hauwiller, X. W. Zhang, W. I. Liang, C. H. Chiu, Q. Zhang, W. J. Zheng, C. Ophus, E. M. Chan, C. Czarnik, M. Pan, F. M. Ross, W. W. Wu, Y. H. Chu, M. Asta, P. W. Voorhees, A. P. Alivisatos and H. M. Zheng, Nano Lett. **18** (10), 6427-6433 (2018).
22. R. L. Sacci, J. M. Black, N. Balke, N. J. Dudney, K. L. More and R. R. Unocic, Nano Lett. **15** (3), 2011-2018 (2015).
23. R. Mishra, R. Ishikawa, A. R. Lupini and S. J. Pennycook, MRS Bull. **42** (9), 644-652 (2017).
24. Z. Q. Yang, L. C. Yin, J. Lee, W. C. Ren, H. M. Cheng, H. Q. Ye, S. T. Pantelides, S. J. Pennycook and M. F. Chisholm, Angew. Chem.-Int. Edit. **53** (34), 8908-8912 (2014).
25. A. Maksov, O. Dyck, K. Wang, K. Xiao, D. B. Geohegan, B. G. Sumpter, R. K. Vasudevan, S. Jesse, S. V. Kalinin and M. Ziatdinov, npj Comput. Mater. **5**, 8 (2019).
26. J. C. M. Toma Susi, Jani Kotakoski, Ultramicroscopy **180**, 163–172 (2017).
27. P. Tan, N. Xu and L. Xu, Nature Physics **10** (1), 73-79 (2014).
28. N. A. Mahynski, E. Pretti, V. K. Shen and J. Mittal, Nat. Commun. **10**, 11 (2019).
29. R. Ignatans, D. Damjanovic and V. Tileli, Physical Review Materials **in press** (2020).
30. T. Matsumoto and M. Okamoto, IEEE Transactions on Ultrasonics, Ferroelectrics, and Frequency Control **57** (10), 2127-2133 (2010).
31. T. Matsumoto, M. Koguchi, K. Suzuki, H. Nishimura, Y. Motoyoshi and N. Wada, Applied Physics Letters **92** (7) (2008).
32. R. Ignatans, D. Damjanovic and V. J. a. p. a. Tileli, (2020).

33. T. Hastie, R. Tibshirani and J. Friedman, *The Elements of Statistical Learning: Data Mining, Inference, and Prediction*. (Springer New York, 2013).
34. R. K. Vasudevan, A. Belianinov, A. G. Gianfrancesco, A. P. Baddorf, A. Tselev, S. V. Kalinin and S. Jesse, Appl. Phys. Lett. **106** (9) (2015).
35. R. K. Vasudevan, A. Tselev, A. P. Baddorf and S. V. Kalinin, ACS Nano **8** (10), 10899-10908 (2014).
36. R. K. Vasudevan, M. Ziatdinov, S. Jesse and S. V. Kalinin, Nano Lett. **16** (9), 5574-5581 (2016).
37. N. Borodinov, W. Y. Tsai, V. V. Korolkov, N. Balke, S. V. Kalinin and O. S. Ovchinnikova, Appl. Phys. Lett. **116** (4), 5 (2020).